\documentclass{osa-article}

\journal{oe}

\articletype{Research Article}

\begin{document}

\title{Widely Tunable, Low Linewidth, and High Power Laser Source using an Electro-Optic Comb and Injection-Locked Slave Laser Array}

\author{J Connor Skehan, Corentin Naveau, Jochen Schroder, Peter Andrekson\authormark{*}}

\address{Photonics Laboratory, Department of Microtechnology and Nanoscience, Chalmers University of Technology, SE - 412 96 Gothenburg, Sweden}

\email{\authormark{*}Peter.Andrekson@Chalmers.SE } 



\begin{abstract}
We propose a simple approach to implement a tunable, high power and narrow linewidth laser source based on a series of highly coherent tones from an electro-optic frequency comb and a set of 3 DFB slave lasers. We experimentally demonstrate approximately 1.25 THz (10 nm) of tuning within the C-Band centered at 192.9 THz (1555 nm). The output power is approximately 100 mW (20 dBm), with a side band suppression ratio greater than 55 dB, and a linewidth below 400 Hz across the full range of tunability. This approach is scalable and may be extended to cover a significantly broader optical spectral range.
\end{abstract}

\section{Introduction}

A single-line, highly tunable, low linewidth, and high power source is highly desireable for applications such as spectroscopy, interferometry, and metrology \cite{fu2017review, newbury2007low, margolis2009frequency, hansen2015quantum}, as well in communications \cite{lu2015research, stern2017compact, qiu2017laser, pilori2019non, gomez201910}. They may also find uses as optical tweezers \cite{norcia2018microscopic}, for the detection of gravitational waves \cite{adhikari2014gravitational}, and more. 

Approaches to achieve this end range from semiconductor distributed feedback lasers (DFB) \cite{duan2018narrow, di2020sub, xu2016narrow, becker2016narrow} and distributed Bragg reflector (DBR) lasers \cite{zimmerman2013narrow, yamoah2019robust, xiang2019ultra, huang2019high, virtanen2016spectral}, to fiber-based ring lasers \cite{fu2017review, shi2014fiber, zhang2016ultra, wu2019simultaneously, ji2012experimental, li2008tunable, al2014ultra}. While these single frequency devices may produce high power and a low linewidth, unfortunately, their tunability is typically low, on the order of a few nanometers. 

To overcome this limitation, we have implemented a system based on a commercial high coherence fiber laser, an electro-optic (EO) frequency comb, and an array of 3 commercially available DFB slave lasers which are used for injection locking of individual comb teeth. The result is a single frequency output which tunable over nearly 1.25 THz within the C-Band ($\sim$10 nm), with approximately 100 mW of power, a side band suppression ratio of more than 55 dB, and less than 400 Hz of linewidth over the source's full range of tunability.

The underlying principle of the system begins with a low linewidth master laser which is modulated to produce cascaded, equally spaced side-bands. These comb lines inherit the optical properties of the source laser, but gain some additional phase noise due to the modulation. This results in the derived spectral linewidth of each tone increasing linearly as a function of the tone's modulation index, $N$. A single tone is then selected using a programmable optical filter (POF) and used for optical injection locking (OIL), a process in which a slave laser's emission is seeded by the injected optical field and overcomes the cavity's own spontaneous emission to induce lasing in accordance with the input optical field. The result is that the slave laser's output ideally follows the phase and frequency of the input field. 

Since the frequency spacing of the EO comb's teeth is equal to the RF driving frequency, then, so long as the source laser is tunable over an optical bandwidth at least equal to the RF driving frequency, the full comb can be shifted to overlap any desired frequency within its bandwidth by red or blue shifting the source frequency. Moreover, since the source laser's tuning range is greater than the frequency spacing of the EO comb, the achievable range may be extended by shifting the full comb beyond its overlap limit. To further increase the achievable frequency range, the approach may be upscaled via the use of additional slave lasers and the use of a broader comb source.

\section{Experimental Details}

\begin{figure}
    \centering
    \includegraphics[width=11cm]{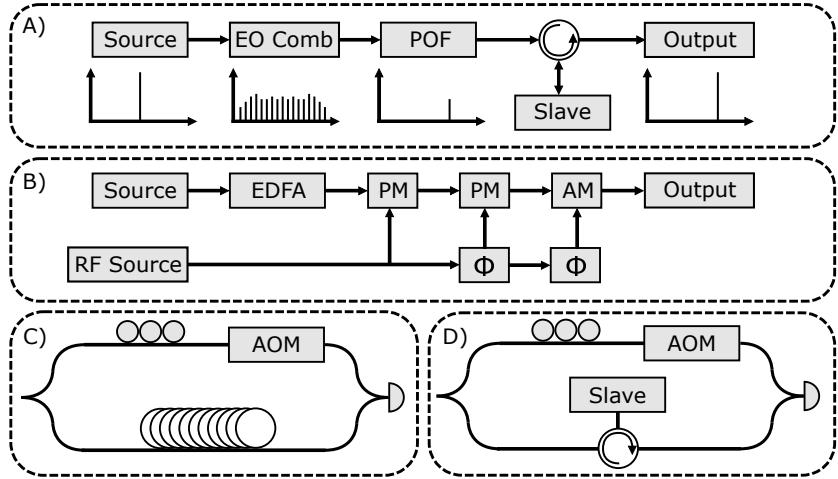}
    \caption{A) The principle of operation, where a source is modulated to produce an EO Comb, a tone is then selected using a programmable optical filter, and a slave laser is injection locked to the selected tone. B) The generation of the optical frequency comb, using a fiber laser as the source, an erbium doped fiber amplifier (EDFA), two phase modulators (PM), one amplitude modulator (AM), some RF tone (in this case, a VCO), and two phase shifters. C) A delay line interferometer. D) A master-slave interferometer.}
    \label{fig1}
\end{figure}

The principle of operation is seen in figure \ref{fig1} A), a fiber laser is used to produce an electro-optic frequency comb, after which a tone is selected using a programmable optical filter and used as the master field for injection locking a subsequent slave laser. 

More specifically, as seen in figure \ref{fig1} B), the comb is created by injecting the source laser into an erbium doped fiber amplifier (EDFA) for amplification, and subsequently into a series of phase modulators (PM) which modulate the source to produce cascaded side-bands \cite{parriaux2020electro, tong2012spectral, slavik2011stable, ishizawa2019ultra,ishizawa2013phase,hisatake2005generation}. These side-bands are equally spaced from each other in accordance with the driving frequency, and inherit the noise properties of the seed laser and the RF source which modulates them. The breadth and shape of the produced optical spectra is represented mathematically as a summation of Bessel functions of the $1^{\textrm{st}}$ kind \cite{parriaux2020electro}.

\begin{equation}
\label{phase-modulated-comb}
    \tilde{A}(\omega) = A_0\sum_{N=-\infty}^{+\infty}J_n(KV_0)e^{iN\phi}\delta(\omega-N\omega_m - \omega_c)
\end{equation}

Here, $A$ is the amplitude, $J_n$ are Bessel functions of the $1^{\textrm{st}}$ kind which act as the envelop function which defines a comb of frequencies $\omega = N\omega_m + \omega_c$ where $N$ is the comb tooth index, $\omega_m$ is the modulation frequency, $\omega_c$ is the fundamental carrier frequency, $K$ is the modulation index, $V_0$ is the amplitude of the modulating signal, and $\phi$ is the static phase noise induced by the voltage feed.

Importantly, the induced phase noise from the driving voltage grows linearly with tooth index, $N$, as per the $e^{iN\phi}$ term in equation (\ref{phase-modulated-comb}). This corresponds to a linear increase in linewidth as the absolute tooth index increases, at a rate which depends on the phase noise of the RF source.

Next, this series of equally spaced frequency tones is injected into an Mach-Zehnder amplitude modulator (AM) which flattens the spectra. Note that the AM is not strictly necessary and could be omitted to extend the tuning range. However, its presence eases the characterization, hence we employed it in this setup.

When the phase offset between the driving tones of the PMs and AMs is correctly aligned, the result is a broad, flat-top EO comb which is sent into a programmable optical filter. This filter selects individual teeth from the comb, either to have their linewidth measured or to act as a master for a subsequent slave laser.

Figure \ref{fig1} C) shows the linewidth measurement system, which is based on the interference of two partially coherent beams. The detection method uses an optimized fiber length of 2.7 km, and electronically measures the peak and trough of the first interference peak, as well as the frequency spacing to the first interference peak. From these data points, we can estimate the linewidth of the laser with high accuracy, as has been previously demonstrated both theoretically and experimentally \cite{wang2020ultra, huang2016laser}. We refer the reader to \cite{wang2020ultra} for a detailed analysis of the method used here. Our length of fiber delay line and choice of detector corresponds to an integration of the phase noise from 75 kHz (as determined by the fiber delay line length) up to 125 MHz (the receiver bandwidth).

Figure \ref{fig1} D) shows a master-slave interferometer. Here, the source is split in two, one arm of which has its polarization modified to maximize interference with the opposite arm, and is frequency offset using an acousto-optic modulator (AOM) to offset frequencies from base-band, while the other arm passes through a circulator and into the slave laser, which in this case is a DFB laser without any optical isolator. By interfering the two beams on a photodiode and measuring the electrical output, we can verify that the slave laser cavity is locked to the master.

\begin{figure}
    \centering
    \includegraphics[width=10cm]{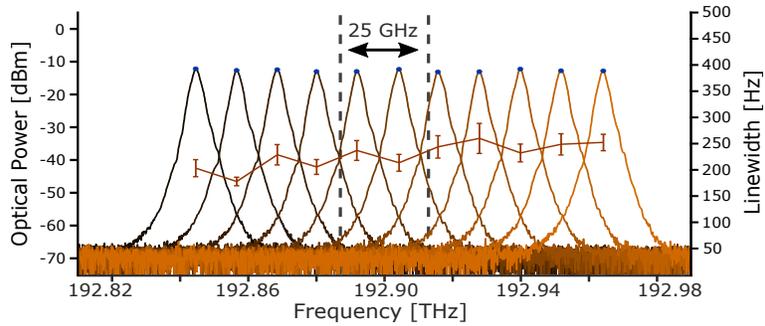}
    \caption{The optical spectrum of the source fiber laser over its full range of tunability, as well as the linewidth measured at each frequency. Optical measurements were taken with an OSA directly after the source with a resolution of 0.1 nm. The double sided arrow and dashed lines indicates a 25 GHz wide band, corresponding to the frequency of RF modulation used for comb creation.}
    \label{fig2}
\end{figure}

Figure \ref{fig2} shows the fiber laser's optical spectra as measured on an optical spectrum analyzer (OSA) and its linewidth as measured using the linewidth characterization system described earlier, over its full tunable frequency range. We can tune over a 115 GHz bandwidth, which corresponds to approximately 0.95 nm. The OSNR stays nearly constant at approximately 55 dB (with 0.1 nm resolution). The linewidth varies between approximately 200 Hz and 250 Hz over the tuning range. The double sided black arrow and dashed lines correspond to the frequency of RF modulation (25 GHz), and therefore the FSR of the EO Comb. Since the fiber laser's tunability is at least equal to the free spectral range (FSR) of the EO comb, the output can be tune to any frequency within the comb bandwidth by tuning the source laser and selection the appropriate comb line. Moreover, because the source tunability is greater than the frequency spacing of the comb (25 GHz), an additional 90 GHz of tunable range is added to the system via tuning the source past the comb's FSR.

\begin{figure}
    \centering
    \includegraphics[width=10cm]{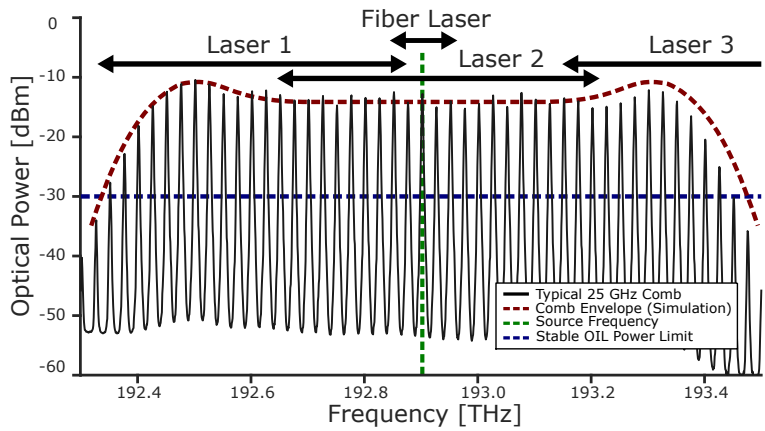}
    \caption{A typical 25 GHz EO frequency comb as produced using the method described above in figure \ref{fig1}. Measurements were taken with an OSA tap directly after comb creation and with a resolution of 0.1 nm. Also indicated is a theoretical spectral envelope as produced via simulation, the central driving frequency of the fiber laser source at 192.9 THz, the stable injection locking limit of -30 dBm, the fiber laser tunability as seen in figure \ref{fig2}, and the available lasing frequencies for the array of DFB slave lasers, whose exact limits are seen in table \ref{tab1}.}
    \label{fig3}
\end{figure}

Figure \ref{fig3} shows a typical EO comb as created via modulation of the fiber laser, as well a comb envelope simulated using equation (\ref{phase-modulated-comb}), the available tuning frequency of the fiber laser the DFB slave lasers (also seen in table \ref{tab1}), and the power limit of stable injection locking. Here, the PM AM phase offsets of the comb are optimized for a balance of bandwidth and flatness, and the DFB slave lasers are tuned via a change in temperature. In general, for the injected beam to overcome the amplified spontaneous emission (ASE) and seed the lasing process \cite{liu2019optical,murakami2003cavity} to induce optical injection locking, a variety of conditions must be met, related to the ratio of power in to fixed output power, linewidth enhancement factor of the cavity, and cavity quality factor \cite{mogensen1985locking, bordonalli1999high}. Here, we fix all variables except the input power to the cavity which we allow to change, and find that at approximately -30 dBm of power, the injection locking is stable over a period of multiple days.

\begin{table}[]
\centering
\begin{tabular}{c|c|c|c}
                 & \textbf{Min. Frequency} & \textbf{Max. Frequency} & \multicolumn{1}{l}{\textbf{Frequency Range}} \\ \hline
\textbf{Laser 1} & 192.330 THz                & 192.875 THz                & 0.545 THz                                    \\ \hline
\textbf{Laser 2} & 192.645 THz                & 193.215 THz                & 0.570 THz                                    \\ \hline
\textbf{Laser 3} & 193.150 THz                & 193.710 THz                & 0.560 THz                                   
\end{tabular}
\caption{The available temperature tuning of each slave laser in the DFB laser array.}
\label{tab1}
\end{table}

\section{Results and Discussion}

\begin{figure}
    \centering
    \includegraphics[width=12cm]{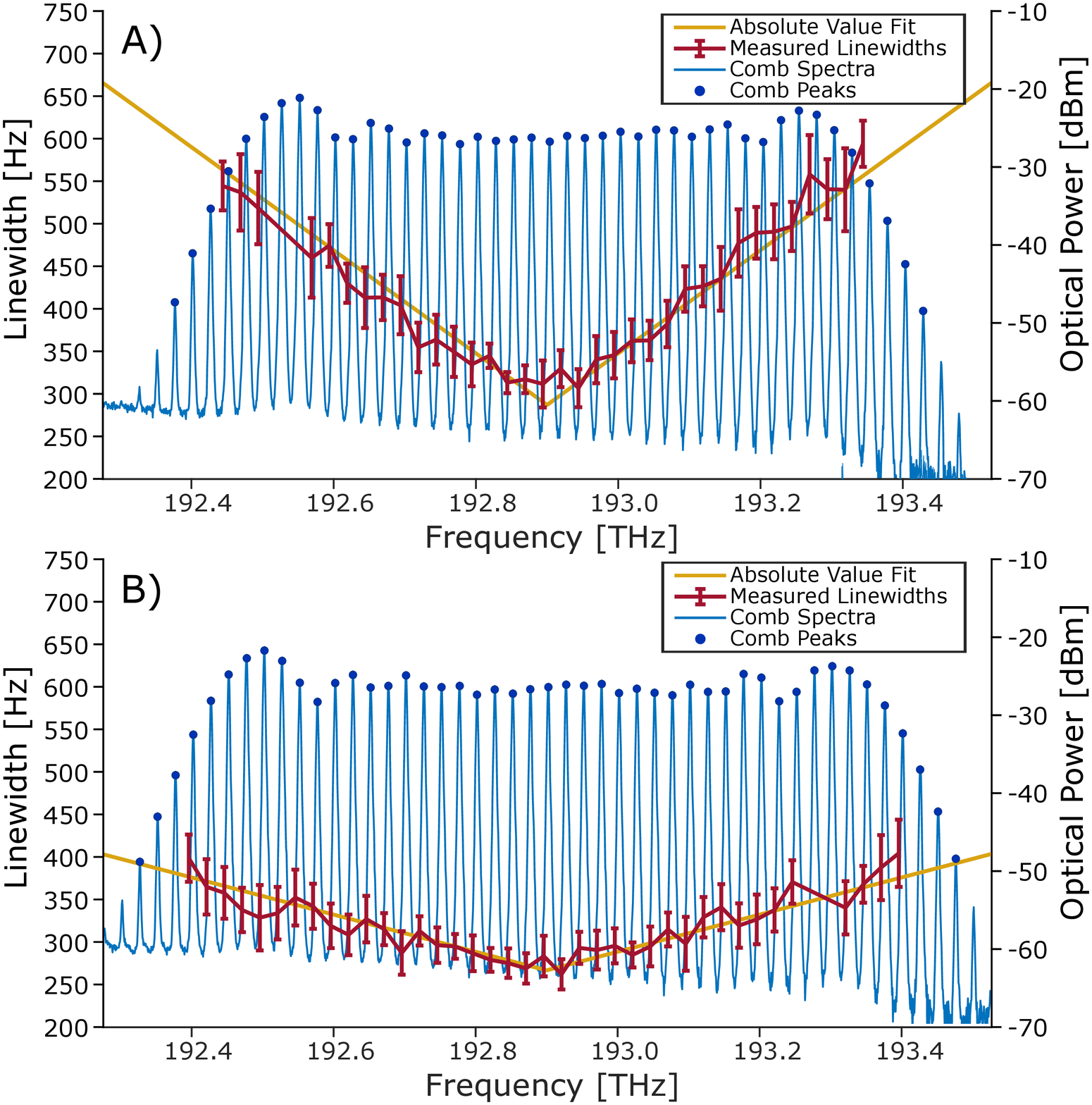}
    \caption{A) Characterization of the unlocked 25 GHz EO comb when modulated using a commercial VCO B) A Characterization of the unlocked 25 GHz EO Comb when modulated using a commercial low-noise RF synthesizer. In all cases, 15 measurements were taken at each frequency, and the error bars represent the standard deviation of these measurements.}
    \label{fig4}
\end{figure}

As seen in figure \ref{fig4}, we begin by fully characterizing the linewidths of each unlocked comb tone as a function of frequency and for two different types of RF modulation. In figure \ref{fig4} A), a small and relatively inexpensive voltage controlled osillator (VCO) is used, and in figure \ref{fig4} B) a bulky, but tunable low-noise RF synthesizer is used. All measurements were taken 15 times with a minimum of -23.5 dBm on the photodiode to stay outside of the power-limited regime, and in all cases the $N = 0$ tone consistently approaches a linewidth value near 275 Hz.

Two full sets of measurements were taken using a commercial VCO, and an absolute value function is fit to the recovered data. On the first day we find a linewidth growth of approximately 606.1 Hz/THz, and two days later we find a linewidth growth of approximately 604.6 Hz/THz. Both values are within the error of fitting. Alternatively, this can be approximated as a linewidth growth of approximately 15 Hz/$N$. The data corresponding to the first day of operation is seen in figure \ref{fig4} A). 

\ref{fig4} B) depicts similar data which corresponds to identical conditions except for a change in RF clock. Here, instead of using a commercial VCO, we use a commercial low-noise RF synthesizer. Predictably, the phase noise is lower, and therefore the linewidth increase as a function of $N$ is reduced. On the first day of measurement when using the RF synthesizer, and for a 25 GHz comb, we found a linewidth growth of approximately 181.9 Hz/THz. Two days later we find a value of 190.1 Hz/THz. Alternatively, this may be represented as a growth of approximately 4.5 Hz/$N$. Data from the first day is pictured.

\begin{figure}
    \centering
    \includegraphics[width=11cm]{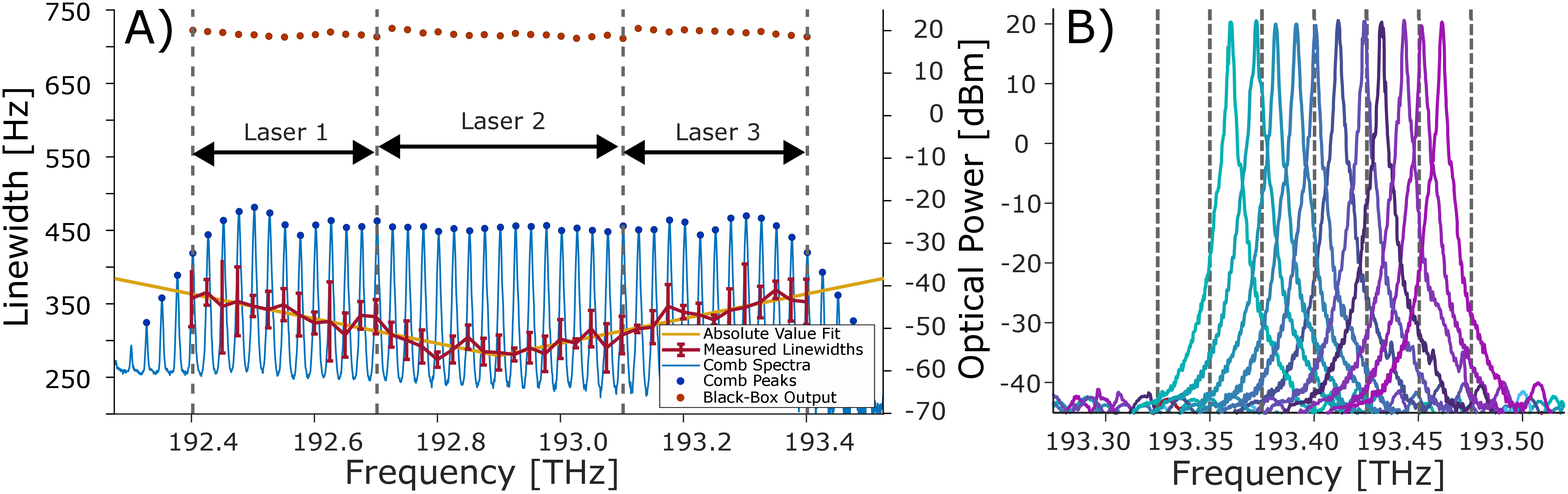}
    \caption{A) Here, each tone of the EO comb has been injection locked to one laser in the DFB array, with the source fiber laser injected at 192.9 THz. Here, all tones with a frequency less than 192.7 THz are locked to Laser 1, all tones above 193.1 THz are locked to Laser 3, and any intermediate tones are locked to Laser 2. In all cases, 15 measurements are taken at each frequency, and the error bars represent the standard deviation of these measurements. Tones below 192.4 THz and above 193.4 THz were not measured due to the instability of injection locking. For all tones, the injection locked power reamins nearly constant across the full spectrum, and 20 dBm (100 mW) of power. B) The output spectrum of the injection locked slave laser, as measured using an OSA at 0.01 nm resolution, after tuning the source fiber laser from 192.95 Thz to 192.85 THz, in increments of approximately 10 GHz. Dashed lines indicate 25 GHz increments.}
    \label{fig5}
\end{figure}

Next, as shown in figure \ref{fig5} A), we individually lock every tone of the 25 GHz EO comb after modulation with the RF synthesizer, and measure the linewidth of the injection locked slave laser output for each $N$ tone. As before, the central tone at 192.9 THz approaches 275 Hz, and the rate of linewidth broadening is 168.2 Hz/THz, or approximately 4.5 Hz/$N$, which is similar to the previously measured value in the unlocked case. In general we observe no significant degradation of linewidth between the unlocked comb teeth in figure \ref{fig4} B) and its individually injection-locked tones as seen in figure \ref{fig5} A), indicating that OIL does not add to the linewidth growth rate as a function of $N$. Of note, the output power of the injection locked spectra is approximately invariant to input power and input frequency, and is fixed at approximately 20 dBm (100 mW) across the full breadth of the comb.  

In figure \ref{fig5} B), we see the output spectrum of of the injection locked $N = 20$ tone as the central $N = 0$ tone (the source laser) is tuned from approximately 192.95 THz to 192.85 THz, as measured on the OSA at 0.01 nm resolution. For all injection locked tones measured, the output power is nearly constant at approximately 20 dBm (100 mW), and the side band suppression ratio is greater than 55 dB. We did not notice any change in the output spectrum between the locked and unlocked state of the DFB lasers at this OSA resolution, indicating that there was little to no degradation of the slave laser's spectrum caused by e.g. ASE from the pre-comb EDFA in the injection locking process.

\begin{figure}
    \centering
    \includegraphics[width=11cm]{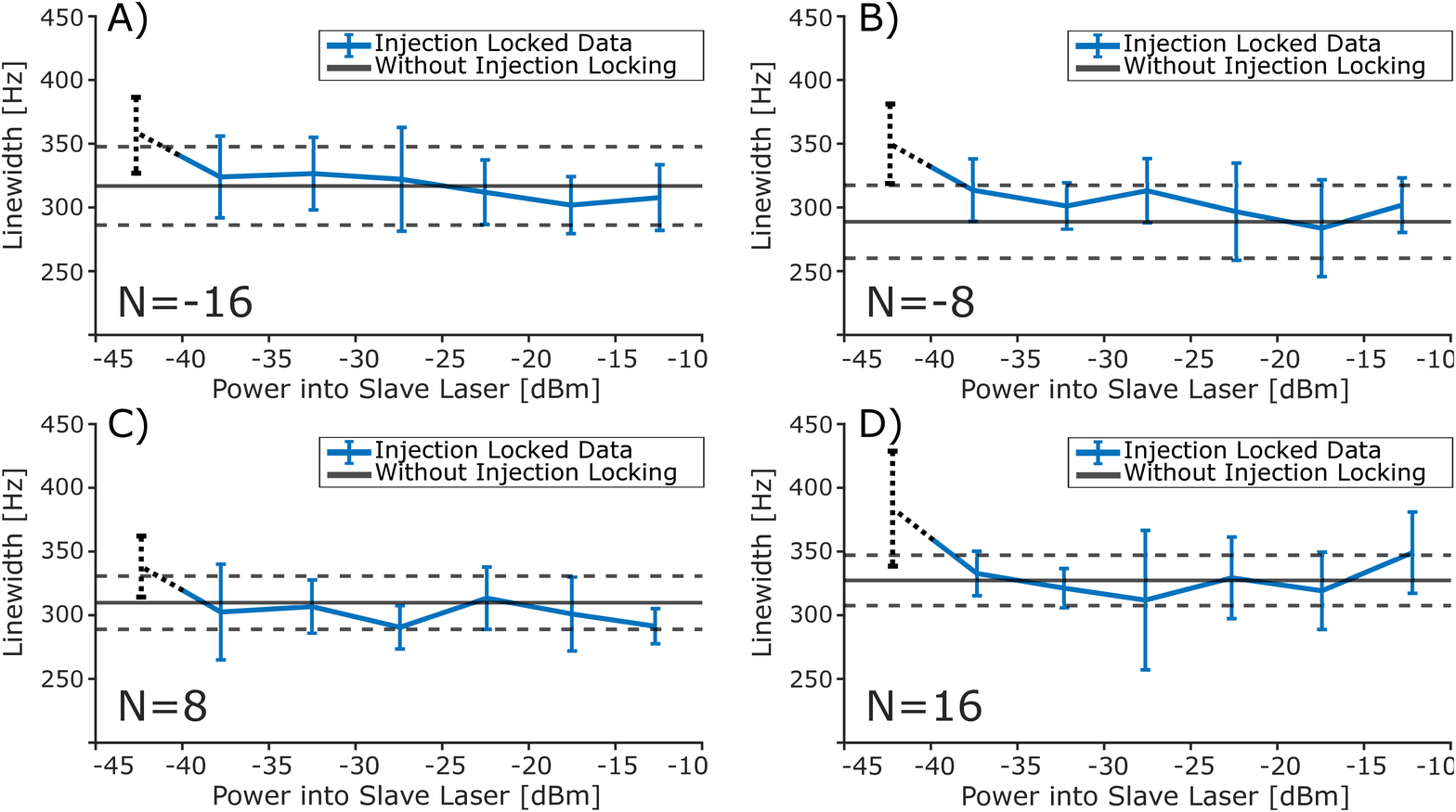}
    \caption{Linewidth versus input power curves for A) $N$ = -16, B) $N$ = -8, C) $N$ = 8, and D) $N$ = 16 comb indices. In all cases, 15 measurements are taken at each frequency, and the error bars represent the standard deviation of these measurements. Here, the dark grey horizontal lines represent the unlocked linewidth, and the horizontal dashed lines represent the error bars of the unlocked linewidth. The dashed black lines indicate the region of unstable OIL.}
    \label{fig6}
\end{figure}

Figure \ref{fig6}), shows the measured linewidth of 4 characteristic tones as a function of injection power. At -30 dBm of power, injection locking is stable over a period of multiple days with no required feedback loops. Below this level, injection locking is only stable over a period of a few hours, and below -40 dBm of input power, staying locked for a period of more than five minutes proved difficult. This ``slipping out of lock'' is what causes the increase in measured linewidth at low power. That said, no significant change in linewidth is observed when locked using between -40 and -12.5 dBm of input power.

\begin{figure}
    \centering
    \includegraphics[width=11cm]{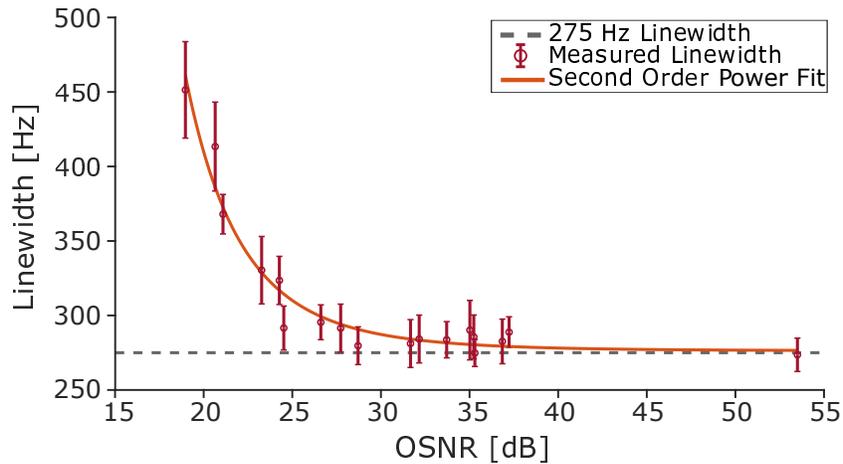}
    \caption{The measured linewidth of the slave laser as a function of input OSNR (0.1 nm bandwidth), at a central pump frequency of 192.9 THz. Here, 10 measurements are taken at each frequency, and the error bars represent the standard deviation of these measurements. All measurements are taken well outside of the power-limited regime.}
    \label{fig7}
\end{figure}

To complete the investigation into the limits of injection locking, the $N = 0$ tone is selected from the comb using a POF, incrementally attenuated, and subsequently amplified up to -30 dBm using an EDFA. The tone is then used as the master for OIL, and the injection locked tone's linewidth is measured. The resulting plot of input OSNR vs measured linewidth is seen in figure \ref{fig7}). Stable optical injection with no observable penalty in linewidth requires approximately 30 dB of OSNR for a fixed input power of -30 dBm \cite{ruiz2019effects}. Even a significant decrease in input OSNR (down to 20 dB) only produces a small increase in measured linewidth, on the order of a few hundred Hz. The use of an additional EDFA in our system therefore permits injection locking with no discernible penalty in linewidth of approximately 6 additional tones, which corresponds to an additional 150 GHz of black-box output tunablity. 

\section{Conclusion and Future Outlook}

In conclusion, we have demonstrated a highly-tunable ($\sim$ 1.25 THz), high power ($\sim$100 mW), high side band suppression ratio ($>$ 55 dB), and low linewidth ($<$ 400 Hz) single frequency laser via the creation of an EO comb, selection of individual comb teeth via a POF, and subsequent optical injection locking of these tones using a set of three DFB slave lasers.

Assuming a 25 GHz comb created using the RF synthesizer as above and a pump at the center of the C-Band (193.725 THz, or 1547.5 nm) whose linewidth is approximately 275 Hz, the full C-Band could be reached with less than 750 Hz of linewidth across its range. Alternatively, the full C-Band could be reached with less than 1.5 kHz of linewidth across its full range using a commercial VCO.

However, in our case the EO comb spectra fails to cover the full C-Band. In the future, this problem may be remedied by up-scaling the system via the insertion of additional phase modulators before the amplitude modulator during the process of comb-creation, via non-linear broadening of the comb in HNLF or photonic waveguides, or the use of multiple frequency combs / source lasers (which may serve to even further reduce the highest observed linewidth in such a system). Alternatively, it has been shown that stable injection locking down to -70 dBm is possible with the use of phase-locked loops \cite{kakarla2018optical}, although it should be noted that this adds significant complexity to the system.

Finally, there may be interest in seeding the EO comb using a source which is even more tunable or even lower in linewidth than the fiber laser used here, in increasing the modulation frequency and thus FSR of the comb, in using a comb source other than an electro-optic comb such as a dark soliton comb \cite{xue2015mode, helgason2021dissipative}, or in replacing the DFB slave laser array with a single Fabry-Perot type slave laser. All solutions may provide various benefits or drawbacks in terms of achievable bandwidth, linewidth, price, complexity, and size. This work represents a first step in the exploration of this parameter space, and an description the general approach.

\section{Acknowledgements}
Funding was provided by the Swedish Research Council (VR) under project grant VR-2015-00535 and by the KA Wallenberg Foundation.

\bibliography{sample}

\end{document}